\documentstyle[12pt]{article}
\def\Journal#1#2#3#4{{#1} { #2} (#3) #4}
\def\CQG{ Class. Quantum Grav.}
\def\GRG{ Gen. Rel. Grav.}

\def\PLA{{ Phys. Lett.}  A}

\def\PRL{ Phys. Rev. Lett.}

\def\RMP{ Rev. Mod. Phys.}

\begin{document}
\begin{center}

{\large\bf General relativistic galvano-gravitomagnetic effect\\
in current carrying conductors}\\
 \vspace*{0.5cm}

{\large\bf B.J. Ahmedov}\footnote{E-mail:
ahmedov@astrin.uzsci.net}\\

\vspace{0.5cm}

{\em Institute of Nuclear Physics and Ulugh Beg Astronomical
Institute\\ Ulughbek, Tashkent 702132, Uzbekistan\\  The Abdus
Salam International Centre for Theoretical Physics\\ 34014
Trieste, Italy}\\
\end{center}
\vspace{1cm}
\begin{abstract}

The analogy between general relativity and electromagnetism suggests that
there is a galvano-gravitomagnetic effect, which is the gravitational
analog of the Hall effect.  This new effect takes place when a current
carrying conductor is placed in a gravitomagnetic field and the
conduction electrons moving inside the conductor are deflected
transversally with respect to the current
flow. In connection with this galvano-gravitomagnetic effect, we explore
the possibility of using current carrying conductors for detecting the
gravitomagnetic field of the Earth.

\end{abstract}

PACS: 04.20.Cv; 04.40.-b; 04.80.Cc

Keywords: General relativity;  Lense-Thirring precession;
Gravitomagnetism;  Ohm's law\\

\newpage

In general relativity, the equation of
motion
for a test particle with mass $m$ and velocity ${\mathbf v}$ in the
weak-field and slow-motion limit has form [1]

\begin{eqnarray}
m\frac{d\mathbf v}{dt}\cong m({\mathbf E} _g+\frac{1}{c}{\mathbf v}
\times{\mathbf B
_g}).  \label{eq:gmot} \end{eqnarray}

Equation (1) shows that there is an interesting and
useful analogy between weak gravitational fields and electromagnetic
fields according to which the gravitational field in the weak limit can be
decomposed
into a ``gravitoelectric" part $${\mathbf E} _g=-\frac{GM}{r^2}{\mathbf
r}$$
and a ``gravitomagnetic" one
 $${\mathbf B}
_g=\frac{2G}{c}[\frac{({\mathbf J}-3({\mathbf J\cdot r}){\mathbf
r})}{r^3}],
%\label{eq:gm}
$$
 with ${\mathbf E}_g$ and ${\mathbf B}_g$ being the analogues of electric
${\mathbf E}$ and magnetic ${\mathbf B}$ fields. $M$ and $\mathbf J$ are
the mass and angular momentum of gravitational object, $\mathbf r$ is the
position of test particle. This analogy brings up the question: are there
gravitational analogues of the effects known in electrodynamics?

Indeed, in the case of low velocities and weak fields, the formal analogy
between gravity and electromagnetism gives rise to a number of similar
phenomena collected under the name gravitoelectromagnetism (see, for
review, [1,2]). Due
to the weakness of the gravitomagnetic effects, however, their existence
has yet to be verified.  For slowly rotating masses, such as the Earth or
the Sun, the gravitomagnetic field is expected to be extremely weak. Even
near the surface of the Earth, the gravitomagnetic contribution is about
$10^6$ times smaller than the gravitoelectric (monopolar) one. However,
this prediction of general relativity has been recently tested [3] by
means of data obtained from
laser-ranging observations of LAGEOS satellites and is expected to be
verified again in the near future via the Gravity Probe B experiment
(which
measures the mechanical precession of superconducting gyroscopes carried
by drag-free satellite in a polar orbit around the Earth [4]).

The use of highly sensitive SQUIDs for detecting general relativistic
effects can significantly reduce the use of mechanical measuring devices
such as gyroscopes and favour the use of equivalent but more precise
electromagnetic devices. This circumstance motivates the investigation of
gravitoelectric and gravitomagnetic effects on electromagnetic processes
in (super-)conductors.

The electric field induced by the gravitoelectric field in conductors has
been widely discussed starting with Shiff and Barnhill (SB) who observed
[5] that electrons inside a metal would sag under gravity, until a
constant gravitoelectric field is balanced by the electrostatic force of
compression. This would create an electric field
$E={mg}/{e}=-5.6\times 10^{-11}V/m$ inside the metal that would
exactly compensate the acceleration due to gravity on the electron
(here $m$ and $e$ are the mass and charge of an electron, $g$ is the
gravitational acceleration).
Moreover, Dessler et al (DMRT) also observed [6] that not only the
electrons but also the ions should sag under gravitoelectric field to
produce an effect $10^3-10^4$ times greater and of oposite sign with
$E\approx +10^{-6}V/m$.  The conclusion is that both effects described by
SB and DMRT occur, but under certain conditions there is a competition
between them [7-9].

Similarly, the influence of the gravitomagnetic field on the magnetic
properties
of (super-)conductors has been investigated by a number of authors
starting with De Witt~\cite{dewitt} and Papini~\cite{papini} both
from an experimental and a theoretical point of view.
For example, proposals have been made to detect the magnetic London
moment generated by the gravitomagnetic field in superconductors even if
the present technology is still insufficient to detect such
weak magnetic fields.

However, we expect that the interplay between the gravitomagnetic field
and the electric current can amplify the general relativistic effects and
the use of it might therefore improve the experimental accuracy. In this
respect, conductors are more suitable for research than superconductors,
since electric currents vanish inside the superconducting media.  To the
best of our knowledge no attempt has been made to investigate such effects
and our goal here is to study the gravitomagnetic
effects on the electric current flowing in a conductor with the aim to
find new general relativistic effects.

It is well-known that the macroscopic consequence of the Lorentz force on
the conduction current is the Hall effect~\cite{hall}, according to which
an electric field

\begin{eqnarray}
{\mathbf E}_H=R_H{\mathbf
j}\times{\mathbf B}
\label{eq:hall}
\end{eqnarray}

appears across a conductor immersed in a magnetic field ${\mathbf B}$ and
crossed by a current ${\mathbf j}$. Here $R_H$ is the Hall constant.
Driven by the gravitomagnetic analogies we might ask whether the
gravitomagnetic field can act on the moving conduction electrons in the
media as does the magnetic one, i.e. is there gravitomagnetic analog of
the Hall effect?

In order to get an answer to this question we first formulate
Ohm's law for a conductor embedded in the weak external gravitational
field derived from the equation of motion of conduction electrons
inside a conductor subjected to an electromagnetic and
gravitational fields
\begin{eqnarray}
m\frac{d{\mathbf v}}{dt}\cong m({\mathbf E}_g+\frac{1}{c}{\mathbf
v}\times{\mathbf B}_g)-e({\mathbf
E}+\frac{1}{c}{\mathbf v}\times{\mathbf B})+\frac{ne^2}{\lambda}{\mathbf
v}.
\label{eq:model}
\end{eqnarray}
The last term in (\ref{eq:model}) is due to the resistance force acting on
conduction
electrons from continuous medium with conductivity $\lambda$,
$n$ is the electron's concentration.

In a steady state,
with $\frac{d {\mathbf v}}{dt}=0$ everywhere,
the equation (\ref{eq:model}) gives Ohm's law
\begin{eqnarray}
{\mathbf E}=\frac{1}{\lambda}{\mathbf j}-R_H{\mathbf j}\times
{\mathbf B}+\frac{A}{c^2}{\mathbf E}_g+
\frac{R_{gg}}{2c} {\mathbf j}\times{\mathbf B}_g,
\label{eq:ohm}
\end{eqnarray}

where ${\mathbf j} =ne{\mathbf v}$ is conduction current density.  Here
$A=\frac{(m-\gamma M_a)c^2}{e}$ and $R_{gg}=\frac{2m}{ne^2}$ are the
coefficients of proportionality, the value of a parameter $\gamma$
depends on model and typically is of order $0.1$ for metals
~\cite{ahm94}.  In the formula (\ref{eq:ohm}) we add the term arising from
the change of the
electrochemical potential $\mu$ in the weak gravitational field and being
proportional to the atomic mass $M_a$ and gravitoelectric field $E_g$
(see, for example,~\cite{{ahm94},{opat88},{anand}}).

The first two terms on the right hand side of equation (\ref{eq:ohm}) are
standard classical terms. The third contribution is due to the
gravitoelectric field. The last term in the equation (\ref{eq:ohm}) is
new and never discussed before. It has pure general relativistic
nature without Newtonian analog, and is caused by the effect of
gravitomagnetic force on the conduction current.

According to expression (\ref{eq:ohm}), an azimuthal voltage
\begin{eqnarray}
V=\frac{R_{gg}}{2cd}i_rB_g
\label{eq:gmvolt}
\end{eqnarray}
will be induced across a conductor carrying radial current $i_r$ even if
an external magnetic field is assumed to be absent ($d$ is
the thickness of the conductor).  Conduction electrons moving in radial
direction would be deflected in azimuthal direction until the
gravitomagnetic force is compensated by the electric field arising from
charge
separation on the lateral sides of the conductor. This new general
relativistic effect arising from
newly incorporated term in Ohm's law is of purely gravitomagnetic origin
and so can be referred to as galvano-gravitomagnetic one
(galvano-gyroscopic
effect, which is the rotational analog of the Hall effect, is discussed
in~\cite{ahm99}).

Alternatively, the equation (\ref{eq:ohm}) can be obtained directly from
the general relativistic constitutive equation ~\cite{ahm98}

\begin{eqnarray}
F_{\alpha\beta}u^{\beta}=\frac{1}{\lambda}\jmath _{\alpha}+
R_H(F_{\nu\alpha}+u_\alpha u^\sigma F_{\nu\sigma})\jmath^\nu
+Aw_\alpha -b\jmath^\beta A_{\alpha\beta}
\label{eq:grohm}
\end{eqnarray}

for conductors embedded in the stationary gravitational field. Here
$F_{\alpha\beta}$ is the electromagnetic field tensor,
$E_\alpha=F_{\alpha\beta}u^\beta$ and
$B_\alpha=-\frac{1}{2}e_{\alpha\beta\mu\nu}u^\beta F^{\mu\nu}$ are the
electric and magnetic fields as measured by an observer at rest with
respect to the conductor, $A_{\beta\alpha}=u_{[\alpha
,\beta]}+u_{[\beta}w_{\alpha]}$ is the relativistic rate of rotation,
$w_\alpha=u_{\alpha;\beta}u^\beta$ is the absolute acceleration,
$b=\frac{2mc}{ne^2}$ is the parameter for the conductor with the four
velocity $u^\alpha$, $[\cdots]$ denotes the antisymmetrization. The last
term in the right hand side of equation (\ref{eq:grohm}) is the one caused
by the gravitomagnetic effect on conduction current.

Ohm's law for conduction current has been generalized to include effects
of gravity and inertia in the recent papers
~\cite{{opat96},{ahm94},{anand}}, but the effect of gravitomagnetic force
upon an electric current has been taken into account only in~\cite{ahm98}.
Recently, Khanna~\cite{khan} has derived the general relativistic Ohm's
law for a two-component plasma and concluded that it has no new terms as
compared with special relativity in the limit of quasi-neutral plasma.
The gravitomagnetic terms did not appear in Ohm's law because of the
magnetohydrodynamic approximation used in~\cite{khan} for the fully
ionized plasma, but would appear in the case of a weakly ionized uncharged
plasma.

Suppose the conductor is at rest in the spacetime of the slow
rotating
gravitational body of mass $M$ and angular momentum
$J$~\cite{mash84}\footnote{ For the Earth, the dimensionless
gravitomagnetic contribution $GJ/c^3r^2 < 4\times 10^{-16}$ is much less
than the gravitoelectric one $GM/c^2r < 7\times 10^{-10}$. Hence, in what
follows the angular momentum will be taken into account only to first
order.}:

\begin{eqnarray}
ds^2=-\Big(1-\frac \alpha r\Big)
(dx^0)^2+\Big(1-\frac \alpha r\Big)^{-1}dr^2+r^2d\theta ^2+r^2\sin
{}^2\theta d\varphi ^2-\nonumber\\
\frac{2a \alpha \sin {}^2\theta }{cr}dx^0d\varphi
\label{eq:lt}
\end{eqnarray}

with $a=J/cM$ and $\alpha =2GM/c^2$. Then with the absolute acceleration
of conductor $$w_\alpha\{0,-\frac{\alpha/2r^2}{1-\alpha/r},0,0\}$$ and
with the nonvanishing components of the relativistic rate of rotation
$$A_{13}=\frac{a \alpha /2r^2}{c(1-\alpha/r)^{3/2}} \sin {}^2\theta ;
\quad A_{23}=-\frac{ a \alpha /r}
{c(1-\alpha/r)^{1/2}}\sin\theta\cos\theta,$$ the
the general-relativistic Ohm's law (\ref{eq:grohm}) takes the form
(\ref{eq:ohm}).

However, metric (\ref{eq:lt}) can be transformed to the reference
frame of a satellite orbiting at a radius $r_0$ with angular velocity
$\Omega =\epsilon\sqrt{\frac{\alpha c^2}
{2r_0^3}}$:
\begin{eqnarray}
ds^2=-\Big(1-\frac \alpha r\Big)
(dx^0)^2+\Big(1-\frac \alpha r\Big)^{-1}dr^2+r^2d\theta ^2+r^2\sin
{}^2\theta d\varphi ^2+\nonumber \\
2(\frac{\epsilon\Omega r^2}{c}-\frac{a \alpha}{cr})
\sin {}^2\theta dx^0d\varphi ,
\label{eq:geod}
\end{eqnarray}
assuming
$+$ and $-$ signs of parameter $\epsilon=\pm$ apply to direct and
retrograte orbits, respectively.

Ohm's law (\ref{eq:grohm}) in the metric (\ref{eq:geod}) takes form
\begin{eqnarray}
{\mathbf E}=\frac{1}{\lambda}{\mathbf j}-R_H{\mathbf j}\times
{\mathbf B}+\frac{A}{c^2}{\mathbf E}_g+
\frac{R_{gg}}{2c}{\mathbf j}\times{\mathbf B}_g
-R_{gg}{\mathbf j}\times{\mathbf\Omega}.
\label{eq:ohmorbit}
\end{eqnarray}

Restricting ourselves to an equatoril orbit, for the sake of simplicity,
we
calculate from (\ref{eq:ohmorbit}) the voltage across a conductor
carrying radial current $i_r$
\begin{eqnarray}
V=V_{Sch}+V_{LT}=
\epsilon\frac{R_{gg}}{d}i_r\Omega+\frac{R_{gg}}{2cd}i_rB_g.
\label{eq:orbitvolt}
\end{eqnarray}

The first term in the right hand side of (\ref{eq:orbitvolt}) results from
the rotational motion of the conductor through the curved space-time
around the Earth and is proportional to the square root of the
Schwarzschild parameter $\alpha$.  We are here interested in the
second term which is due to rotation of central body and corresponds
to the gravitomagnetic effect. For experimental purposes, it is clearly
better to eliminate the lower order terms in expression
(\ref{eq:orbitvolt}).
In order to do this, we could send two conductors in geosynchronous
and
anti-geosynchronous orbits and compare their data when they pass
overhead. The
gravitomagnetic term will have the same sign for both satellites but
Schwarzschild
term changes the sign from one station to another. So the simple addition
of  the data from the counter-rotating stations will isolate the pure
gravitomafnetic term, and therefore, it is
possible to separate gravitomagnetic effect from the dominant
Schwarzschild effect.

The gravitomagnetic effect  may be regarded, among other
things, as a consequence of gravitational
analogue of Larmor's theorem~\cite{mash93} according to which the
effect of the gravitomagnetic force  on an electron is the same as that of
the Lorentz force ${\mathbf F}=-\frac{e}{c}({\mathbf v}\times{\mathbf
B})$
due to magnetic field ${\mathbf B}$, if ${\mathbf B}$ satisfies
\begin{eqnarray}
{\mathbf B}=-(2mc/e){\mathbf\Omega}_{LT},
\label{eq:larmor}
\end{eqnarray}
where the gravitomagnetic field ${\mathbf B}_g$ is identified with the
Lense-Thirring frequency $-2c{\mathbf\Omega}_{LT}$ of the local inertial
frames with
respect to observers at infinity.  Since the
macroscopic consequence of the Lorentz forces on the conduction electrons
is the Hall effect (\ref{eq:hall}), the result (\ref{eq:gmvolt}) follows,
 in fact, from the gravitomagnetic Larmor's theorem (\ref{eq:larmor}).

The value of the potential difference in (\ref{eq:gmvolt})
can be adjusted
by the magnitude of the electric current and this dependence may give one
more possibility to detect the gravitomagnetism by measuring the
difference in a voltage generated across two current carrying conductors
orbiting in pro- and retrograte directions around the Earth.
Of course, ground based laboratory experiment for measurement of the
galvano-gravitomagnetic effect could be also derived. But on Earth, the
angular velocity
of the conductor with respect to a local inertial frame
${\mathbf\Omega}_{L}$ is
given by~\cite{jiu}:

$${\mathbf\Omega}_{L}={\mathbf\Omega}_{VLBI}-{\mathbf\Omega}_{Th}-
{\mathbf\Omega}_{S}-{\mathbf\Omega}_{LT},$$

where ${\mathbf\Omega}_{VLBI}$ is the angular velocity of the laboratory
with
respect to an asymptotic inertial frame, ${\mathbf\Omega}_{Th}$ and
${\mathbf\Omega}_{S}$ are, respectively,
the contributions of the Thomas precession arising from non-gravitational
forces and of the de Sitter or geodetic precession. As a result, in
order to detect
${\mathbf\Omega}_{LT}$ one should measure ${\mathbf\Omega}_{L}$ and
then
substract from it the independently measured value of ${\mathbf\Omega}_{VLBI}$
with VLBI (Very Long Baseline Interferometry, see, for example,
~\cite{koval}) and the
contributions
due to the Thomas and de Sitter precessions. A number of disturbing
effects such as seismic accelerations, local gravitational noise,
atmospheric turbulence have long made ground-based experiment
difficult. Satellite based experiments can offer the
possibility of reduction in these disturbances, although they
will also raise other additional difficulties.

The valuable galvano-gravitomagnetic coefficient $R_{gg}\approx 0.8\times
10^{-22}s$ is estimated for the typical semiconductor. Then the difference
in voltage developed in two counter-rotating semiconductors is
$2V_{LT}=2(R_{gg}/d)i_r\Omega_{LT}\approx 0.6\times 10^{-19}V$ if we put
$a_\oplus=\frac{2}{5} \Omega_\oplus R_\oplus^2$, $\alpha_\oplus =2\times
G\frac{M_\oplus}{c^2}=2\times 0.44cm$, $R_\oplus\approx 6.37\times
10^8cm$, $\Omega_\oplus=7.27\times 10^{-5}rad/s$, $i_r=10^3A$ and
$d=10^{-2}cm$.

In principle, it is possible to measure a voltage of $10^{-19}V$
with today's SQUID tecnology with precise voltage
accuracies~\cite{jain} of $1$ part in $10^{22}V$. In a possible
experiment a superconducting loop of SQUID can be connected across
carrying radial current semiconductor which will form normal layer of
Josephson junction with source of constant voltage $V_{LT}$.
The nonvanishing potential difference $V_{LT}$ would lead to a time
varying magnetic flux through the loop. The change in magnetic flux
$\Phi_b$ inside the circuit during the time interval $[0,t]$ is
\begin{eqnarray}
\Delta\Phi_b =\Delta n\Phi_0+ {c}\int_{0}^{t}V_{LT}dt
\label{eq:flux}
\end{eqnarray}
where $\Phi_0=\pi\hbar c/e=2\times 10^{-7}Gauss\cdot
cm^2$ is quantum of the magnetic flux.
 As long as $\Delta\Phi_b<\Phi_0$,
$n$ will remain constant and $\Delta\Phi_b$ will increase linearly with
time until $\Delta\Phi_b=\Phi_0$, then the order of the step $n$ will
change as flux quantum enters the loop.
Thus this particular loop is sensitive to the $V_{LT}$ and in this
connection to the Lense-Thirring frequency.
Moreover, it should be underlined that the proposed scheme of flux
measurement
(\ref{eq:flux}) is cumulative and the long time measurement can be used
for accumulation of data towards better detection of the gravitomagnetic
effect.

The experiment is quite difficult since the measured effect is small
compared to potential disturbances and careful attention must be paid to
possible systematic experimental errors. While we do not intend to present
here a detailed proposal but to
point out the possibility of new electromagnetic test of general
relativity.

We can foresee two major experimental difficulties: (i) ensuring
that the deviation of the semiconductor's orbit from idealized one is in
the limit of the required accuracy, and (ii) shielding the Earth's
magnetic fields which can induce much bigger voltage due to the Hall
effect and superconducting shells provide a natural means of shielding of
apparatus from external magnetic field. Any experimental project would
need to take these problems into serious consideration.

In this letter, we have shown that the effect of the gravitomagnetic force
on the conduction current is to induce a galvano-gravitomagnetic potential
difference just as the action of magnetic field is to develop a Hall
voltage across current carrying conductor. This is the general
relativistic analog of the Hall effect and when current carrying
semiconductor is subject to general-relativistic gravitational field of
Earth, a potential difference around $10^{-19}V$ will be developed across
it. The general relativistic galvano-gravitomagnetic effect seems to be
experimentally verifiable with the present technology. Because of the
great importance of gravitomagnetic fields for astrophysics and
fundamental physics, such experiment would also constitute an important
and direct measurement of the general relativistic Lense-Thirring effect.

 \section*{Acknowledgements}

The main part of work was carried out when the author was visiting
the Abdus Salam International Centre for Theoretical Physics,
Trieste, during the autumn of 1998.  He is grateful to the
AS-ICTP, for the generous financial support which made this visit
possible and thanks Luciano Rezzolla for his invaluable work on
the text and helpful comments. The research is also supported in
part by the UzFFR (project 01-06) and projects F.2.1.09, F2.2.06
and A13-226 of the UzCST.

%\section*{References}

\end{document}